\documentstyle[aps, manuscript, epsf]{revtex}

\title{Thermodynamical fluctuations and photo-thermal shot noise\\
	in gravitational wave antennae}

\author{V. B.\ Braginsky,\ M.L.\ Gorodetsky and S. P.\ Vyatchanin}

\address{Physics Faculty, Moscow State University, \\
	Moscow 119899, Russia \\
	e-mail: brag@hbar.phys.msu.su}

\begin {document}

\maketitle

\begin{abstract}
Thermodynamical fluctuations of temperature in mirrors of gravitational wave
antennae are transformed through thermal expansion coefficient into
additional noise. This source of noise, which may also be interpreted as
fluctuations due to thermoelastic damping, may not be neglected and leads
to the necessity to reexamine the choice of materials for the mirrors.
Additional source of noise are fluctuations of the mirrors' surfaces
caused by optical power absorbed in dielectrical reflective layers.
\end{abstract}

\section{Introduction}

The full scale terrestrial gravitational wave antennae are now in the
process of assembling and tuning. It is expected that in the year 2001
in one of these antennae (project LIGO-I) the sensitivity in units
of amplitude of the perturbation of metric will be at the level
$h\simeq 1\times 10^{-21}$ \cite{abr1,abr2}. This value means that the optical
readout system will be able to register small displacement between two
test masses (two heavy optical mirrors placed in vacuum pipe at the distance
$L=4\times 10^5$ cm) with the resolution $\Delta L_{\rm grav}
\simeq \frac{1}{2}\, h L\simeq 2\times 10^{-16}$~cm. This displacement may be
produced by a burst of gravitational wave generated in
astrophysical catastrophe. The characteristic frequency in this wave packet
and, correspondingly, in the displacement $\omega_{\rm grav}$ has to be of the
order of $2\pi\times 10^2$~s$^{-1}$. The improvement of the readout system and
better isolation of the mirrors from different sources of noise will permit
to achieve higher sensitivity: in the year 2005 the planned level of
sensitivity has to be better than $h\simeq 2\times 10^{-23}$ and
correspondingly it will be necessary to register
$\Delta L_{\rm grav}\le 4\times 10^{-18}$~cm. It is worth noting that in the
prototype LIGO (where $ L=4\times 10^3$~cm) the already achieved resolution
was  $ \Delta L_{\rm grav}=3 \times 10^{-16}$~cm \cite{abr2}.

The important feature of this antenna is that the optical readout system does
not register displacements between the centers of the mirrors' masses
but between the internal surfaces of the mirrors. These surfaces are coated
by several thin dielectrical layers, which provide high optical reflectivity
of the mirrors and high sensitivity to the displacements. The necessity to measure the displacements
between the internal surfaces inevitably determines the requirement for
high quality factors $Q_{\rm int }$ (small loss angle) of internal mechanical
modes of the mirrors, which provide relatively small values of random thermal
fluctuations of the mirrors related to the mirrors' center of mass because
the value of $\omega_{grav}$ is substantially smaller than the eigen-frequencies
of internal modes. We shall call as Brownian the fluctuations which
are usually calculated from phenomenological model of structural losses in the
material in order to distinguish them from another type of thermodynamical
fluctuations which we describe below. This requirement for high $Q_{int}$ determines
the choice of the mirrors'  material. On the first stage
of LIGO the mirrors are made of very pure fused silica ($SiO_2$), because the
values of $Q_{\rm int \, SiO_2}$ exceed $2\times 10^7$ \cite{donch,raab}.

At present the possibility to make the mirrors for the second stage of LIGO
from sapphire monocrystal ($Al_2O_3$) is widely debated, because more than
20 years ago the achieved value of $Q_{\rm int\, Al_2O_3}$ for this
material was $3\times 10^8$ \cite{mitr}.

We show below that the value of $Q_{\rm int}$ is not the only parameter
which determines the choice of material from which the mirror has to be
manufactured. Till now the experimentalists' interest has neen concentrated on
the reduction of the Brownian motion of the mirrors' surfaces (as it was
described above), to the reduction of thermal fluctuations of the centers of
mirrors' masses due to the friction in the mirrors' suspension \cite{mitr2,houg}
and to excess noise in the suspensions \cite{ageev}. To our best knowledge
nobody has so far paid attention to nonlinear effects which may be ``responsible''
for additional fluctuations of the mirrors surface, which may exceed traditional
Brownian noise for low frequencies. The simplest of these effects is
thermal expansion, originated from anharmonisity of interaction potential of
atoms in solid state. The following numerical example illustrates the role of
nonzero value of thermal expansion coefficient $\alpha$ of the mirrors
material. Suppose that the surface layer of the mirror having thickness
$l_T\simeq 10^{-2}$~cm  (it is the characteristic length
of the decay of Fourier thermal wave at frequency $2\pi\times 10^2{\rm s}^{-1}$
in fused silica) is heated due to some process on $\delta T$.
Then the distance between the surface of the mirror and its center of mass
will also change on the value of
$\delta l$:
\begin{equation}
\delta l=\alpha\, l\, \delta T\simeq 1\times 10^{-17}\ {\rm cm}\
	\times\left(\frac{\alpha}{5\times 10^{-7}\ {\rm K}^{-1}} \right)
	\times\left(\frac{l_T}{1\times 10^{-2}\ {\rm cm} }\right)
	\times\left(\frac{\delta T}{2\times 10^{-9}\ {\rm K} }\right).
\label{deltal}
\end{equation}

The estimate (\ref{deltal}) shows, that if the mirrors are made of fused
silica  and if the temperature of the first layer ($10^{-2}$~cm thick)
of the mirror changes by 2 nK during $5\times 10^{-3}$~s then the
antenna will register a "burst" of gravitational radiation with amplitude of
the order of $h\simeq 10^{-22}$ with $\omega_{\rm grav}=2\pi\times 10^2$~s.

The aim of this article is to present the results of the analysis of the two
effects which are "responsible" for random fluctuations of temperature
in the bulk of the mirrors and which may mimic at certain level the
gravitational wave if $\alpha \ne 0$: photo-thermal shot noise (due to random
absorption of optical photons in the surface layer of the mirror) and
thermodynamical fluctuations of temperature.

As in both effects we are interested in the fluctuations of the coordinate
averaged over the spot on mirror surface with radius $r_0\simeq 1.5$~cm of
laser beam, which is much smaller than the radius of mirror $\simeq 15$~cm,
we replace  the mirror by half-space: $0\le x <\infty,\ \infty< y <\infty,\ \infty< z <\infty$.
We consider also for simplicity this half-space to be isotropic (the anisotropy
of different constants for sapphire is not very large). The role of these
two effects for the condition to reach the standard quantum limit of
frequency stability was analyzed more than 20 years ago \cite{stability}.

As at room temperature the power radiated from the surface
(Stephan-Boltzmann law) is much lower than the heat exchange due to
thermal conductivity (Fourier law), below we use the simplified
boundary condition:
\begin{equation}
\label{boundaryTC}
 \left. {\partial u(x,y,z,t)\over \partial x}\right|_{x=0}=0,
\end{equation}
where $u$ is the deviation of temperature from the mean value $T$.

For the calculations we use approximation in which we neglect the effects
of limited speed of sound but take into account thermal
relaxation.

\section{Photo-thermal shot noise in the mirrors of the antennae}

The multilayer coating of the mirrors absorbs a fraction of
power $W$ of the optical beam, circulating between the mirrors
of Fabry-Perot resonator. This fraction for the best existing today
coatings has not been measured with good precision yet, but for estimates the
usually accepted value of this fraction is approximately equal to
$5\times 10^{-7}\div10^{-6}$ \cite{kimble}. The value of $W$ sufficient to
reach $h\simeq 10^{-22}$ (which is close to Standard Quantum
Limit (SQL) of sensitivity) is approximately $W \simeq 1\times 10^{6} {\rm
Watt}= 10^{13} {\rm erg/sec}$ or $\dot N_0\simeq 5\times10^{24}$~photon/s
\cite{thorne}. It is reasonable to estimate that each mirror absorbs
$W_{abs}\simeq 1\ {\rm Watt}= 10^7$~erg/s or
$\dot N\simeq 5\times 10^{18}$~photon/s. Each absorbed optical photon having
energy $\hbar \omega_0\simeq 2\times 10^{-12}$~erg gives birth to a bunch
of approximately 50 thermal phonons. Each of these phonons has a
relatively short free path $l^*$ (for $SiO_2$ and $Al_2O_3$ the values are
$l^*_{SiO_2}\simeq 8\times 10^{-8}$~cm and
$l^*_{Al_2O_3}\simeq 5\times 10^{-7}$~cm \cite{path}). After a very short time
interval which is a few times longer than $l^*/v_s\simeq 10^{-12}$~s
(where $v_s$ is the speed of sound) these phonons will produce a local
jump of temperature inside the coating $(5 \div 10)\times
10^{-4}$~cm thick. As we want to know the variance of temperature during
the averaging time $\pi/\omega_{\rm grav}\sim 5\times 10^{-3}$~s which is
several orders longer than $l^*/v_s$ and as during this time a large number
($\dot N \pi/\omega_{\rm grav}\sim 2.5\times 10^{16}$) of photons are
creating bunches of thermal phonons --- it is reasonable to use the model of
shot noise. As the length $l_{T}$ is much smaller than the radius of laser beam
$r_0\simeq 1.5 $~cm for semi-qualitative analysis one can use
one dimensional model, described by thermal conductivity equation:
\begin{eqnarray}
\label{TC1D}
\frac {\partial u(x,t)}{\partial t} -
	a^2 \frac {\partial^2 u(x,t)}{\partial x^2}
        &=&  \frac{w}{\rho C \pi r_0^2}\, 2\delta (x),\\
\label{boundary1D}
\left.{\partial u(x,t)\over \partial x}\right|_{x=0}&=& 0,
                \quad 0\le x < \infty, \\
\langle w(t)\, w(t') \rangle&=& \hbar \omega_0 W_0 \delta(t-t'),\quad
	w(t)=W_{\rm abs} -W_0,	\quad
	W_0= \langle W_{abs}(t)\rangle \nonumber,
\end{eqnarray}
where $a^2=\lambda^*/(\rho C)$, $\lambda^*$ is thermal conductivity, $\rho$
is density and $C$ is specific heat capacity.

Assuming that displacement $X_{\rm 1D}$ (which is of interest) is
proportional to temperature $u$ averaged along the axis $x$
\begin{eqnarray}
X_{\rm 1D}&=& \alpha \int_0^{\infty}dx\ u(x,t).
\end{eqnarray}
it is not difficult to obtain the spectral density of $X_{\rm 1D}$
\footnote{
We use ``one-sided'' spectral density, defined only for positive
frequencies, which may be calculated from correlation function
$\langle  X(t)X(t+\tau)\rangle$ using formula
$$
S_X(\omega)=2\int_{-\infty}^{\infty}d\tau\,
        \langle  X(t)X(t+\tau)\rangle\,\cos (\omega\tau).
$$}:

\begin{eqnarray}
\label{SDTS1D}
S_{\rm TS1D}(\omega)&=& 2\alpha^2\,
	\frac{ \hbar \omega_0 W_0}{(\rho\, C\, \pi r_0^2)^2}\, \frac{1}{ \omega^2}.
\end{eqnarray}

For the exact calculations for half-space (see Appendix A)
we introduce  the displacement $\bar X$ averaged over the beam spot on the surface
\begin{equation}
\label{barX}
\bar X = \frac{1}{\pi r_0^2}\int\!\int_{-\infty}^{\infty} dy\, dz,
        v_x(x=0,y,z)\, e^{-(y^2+z^2)/r_0^2},
\end{equation}
where $v_x(x=0,y,z)$ is $x$-component of vector of deformation on the surface, and find
the spectral density of $\bar X$:
\begin{eqnarray}
\label{SDSN}
S_{\rm TS}(\omega)&=& 2\alpha^2(1+\sigma)^2\,
        \frac{\hbar \omega_0 W_0}{(\rho\, C\,\pi r_0^2)^2}\,
        \frac{1}{\omega^2},
\end{eqnarray}
where $\sigma$ is the Poisson coefficient.

Here and below we use the fact
that for materials at room temperature the thermal relaxation time of the
spot $\tau_T\simeq r_0^2/a^2 $ is very large:
\begin{equation}
\label{condition}
\omega\gg \frac{a^2}{r_0^2}.
\end{equation}

It is easy to see that the exact solution (\ref{SDSN}) differs
from the approximate one (\ref{SDTS1D}) only by Poisson coefficient.

\section{Thermodynamical fluctuations of temperature, thermoelastic damping
and surface fluctuations}

It is known from thermodynamics \cite{Landstat} that the total variance of
fluctuations of temperature in volume $ V$ is described by the
following simple formula:
\begin{eqnarray}
\langle\,\delta T^2\,\rangle =\frac{\kappa T^2}{\rho C V},
\label{tvar}
\end{eqnarray}
where $\kappa$ is the Boltzmann constant.
In classical solid state thermodynamics these fluctuations
are not correlated with fluctuations of volume (which are responsible for
the Brownian noise) \cite{Landstat} but this is not the case for nonzero
thermal expansion coefficient. To find the influence of the fraction of this
type of fluctuations in the vicinity of $\omega_{grav}$
on the vibration of the surface along the x-axis in direct way we use the
Langevin approach and introduce fluctuational thermal sources
$F(\vec r,t)$ added to the right part of the equation of
thermal conductivity:
\begin{eqnarray}
\label{TDFT}
\frac{\partial u}{\partial t} - a^2\Delta u = F(\vec r, t).
\end{eqnarray}

These sources should be normalized in a way to satisfy formula
(\ref{tvar}). By substituting the solution of this equation for the
temperature $u$ into the equation of elasticity and averaging displacement
(\ref{barX})
over the spot on the surface, we find spectral density of its fluctuations
in frequency range of interest (see details in Appendix B):
\begin{eqnarray}
\label{SDTD}
S_{\rm TD}(\omega)&\simeq&\frac{8}{\sqrt{2\pi}}\, \alpha^2(1+\sigma)^2\,
	\frac{ \kappa T^2 }{\rho C}\,
	\frac{a^2}{r_0^3}\, \frac{1}{\omega^2} \qquad \mbox{if}\quad
        \frac{r_0^2 \omega}{a^2}\gg 1.
\end{eqnarray}

We can also find the spectral density of displacement using
Fluctuation-Dissipation theorem (FDT) if we assume that the only dissipation
mechanism in the mirror is thermoelastic damping. Elastic deformations in
a solid body through thermal expansion lead to inhomogeneous  
distribution of temperature and hence to fluxes of heat and losses of energy
\cite{zener}. The same approach was used  \cite{levin,french}
for the calculation of vibration of the surface due to the Brownian fluctuations.
To calculate the fluctuations in this approach we should apply a periodic
pressure $p$ distributed over the beam spot on the surface:
\begin{equation}
p(y,z,t)=\frac{F_0}{\pi r_0^2}\, e^{-(y^2+z^2)/r_0^2}\,e^{i\omega t}
\end{equation}
and we should find susceptibility $\chi =\bar X/F_0$. After that in accordance with FDT
the spectral density of displacement $\bar X$ must be proportional to
imaginary part of susceptibility $\chi $:
\begin{eqnarray}
S_x(\omega)=\frac{4 \kappa T}{\omega} |{\rm Im} (\chi (\omega))|.
\label{spectrum}
\end{eqnarray}

Calculations (see Appendix C) show that FDT gives the same result as formula
(\ref{SDTD}). Therefore, we can conclude that thermodynamical fluctuations of
temperature are the physical source of fluctuations deduced from FDT based on
thermoelastic damping.

Calculations of ${\rm Im} (\chi )$ are rather bulky but Zener \cite{zener} suggested a
very simple approach allowing to estimate
thermoelastic loss angles and hence the imaginary parts of elastic coefficients. It
allows to find  ${\rm Im} (\chi)$ only from formula for $\chi $ in zero order
approximation (see Appendix C):
\begin{eqnarray}
  \chi ^{(0)}(\omega)=\frac{(1-\sigma^2)\, }{\sqrt{2\pi}\, E\, r_0},
\label{rigidity}
\end{eqnarray}
where $E$ is the Young's modulus.

The imaginary parts of elastic coefficients due to thermoelastic damping
may be calculated in Zener's approach according to the formulas \cite{landau}:
\begin{eqnarray}
  \label{zeners}
  \frac{{\rm Im} (\sigma)}{\sigma}=\frac{\sigma_{S}-\sigma_{T}}{\sigma_{T}}\,
        \frac{\omega \tau_{T}}{1+\omega^2\tau^2_{T}}=
        \frac{\alpha^2\,T E }{C\rho}\,
        \frac{1+\sigma}{\sigma}\,
        \frac{\omega \tau_{T}}{1+\omega^2\tau^2_{T}},\nonumber\\
  \frac{{\rm Im} (E)}{E}=\frac{E_{S}-E_{T}}{E_{T}}\,
        \frac{\omega \tau_{T}}{1+\omega^2\tau^2_{T}}=
        \frac{\alpha^2\, T E}{C\rho}
        \frac{\omega \tau_{T}}{1+\omega^2\tau^2_{T}} .
\end{eqnarray}
Where $E_T, E_S $ are consequently iso-thermic and adiabatic Young's moduluses
($E_T\simeq E_S \ \simeq E$) and $\sigma_T,\sigma_S $
are iso-thermic and adiabatic Poisson coefficients
($\sigma_T\simeq\sigma_S  \simeq \sigma$),
$\tau_T$ is the time of thermal relaxation of our spot.
By substituting these values in (\ref{rigidity}) and then in
(\ref{spectrum}) and taking into account that
$\tau_T=2r_0^2/a^2\gg 1/\omega$ (this
relaxation time may be found relatively easy from the solution of the
problem of
thermal relaxation of the heated spot) we finally obtain the same formula
(\ref{SDTD}).

It is really worth noting that all the three methods give precisely the same result.

\section{Numerical estimates}

Now we want to compare the fluctuations of the mirrors' surface caused
by the described above mechanisms with other known sources of noise in LIGO
antennae.

The main determining source of noise associated with the mirrors' material
is usually described by the losses in the model of structural damping (we
denote it as Brownian motion of the surface). In this model the angle of
losses $\phi$ does not depend on frequency and analogously
to \cite{saulson,levin,french} by substituting $E=\bar E(1+i\phi)$ in
(\ref{rigidity}) and then in
(\ref{spectrum}) we obtain:
\begin{eqnarray}
\label{SDSD}
S_{\rm SD}(\omega)&\simeq&\frac {4\kappa T}{\omega}\,
	\frac{(1-\sigma^2)}{\sqrt{2\pi}  E r_o}\,\phi.
\end{eqnarray}
For simplicity here we neglected the unknown imaginary part of Poisson
coefficient.

The sensitivity of gravitational wave antenna to the perturbation of metric
may be recalculated from noise spectral density of displacement $X$
using the following formula:
\begin{equation}
h(\omega)=\frac{\sqrt{2 S_X}}{L},
\label{hsd}
\end{equation}
where we used the fact that antenna has two arms (with length $L$) and in each arm only one
mirror adds to noise (on end mirrors the fluctuations are
averaged over the larger beam spot).

The LIGO-II antenna will reach the level of SQL, so we also
compare the noise limited sensitivity to this limit in spectral form \cite{thorne}:
\begin{equation}
h(\omega)=\sqrt{\frac{8\hbar}{m\omega^2L^2}}.
\label{hsql}
\end{equation}

On figures 1 and 2 we plot the sensitivity limitations for the perturbation of metric
caused by the described thermal mechanisms and SQL (formulas \ref{SDSN},\ref{SDTD},\ref{SDSD},\ref{hsd},\ref{hsql})
for fused silica (Fig.1) and sapphire (Fig.2) mirrors. We used the following
numerical values for the parameters of the two materials --- fused silica and sapphire.
(We should note that the actual figures, which various sapphire manufactures
provide, significantly vary -- tens of percents, especially for
the thermal expansion coefficient):

\begin{eqnarray}
\omega&=&2\pi\times 100\ \mbox{s}^{-1},\quad
	r_0=1.56\  \mbox{cm},\quad
	T=300\ \mbox{K}, \nonumber\\
\omega_0&=&2\times 10^{15}\ \mbox{s}^{-1},\quad
	L=4\times 10^5\ \mbox{cm}, \quad W_0=10^7{\rm erg/s}\label{parameter}; \\
\mbox{Fused silica:}
	\quad \alpha&=&5.5\times  10^{-7}\ \mbox{K}^{-1}, \ \
	\lambda^*=1.4\times  10^5\ \frac{\mbox{erg}}{\mbox{cm s K}},
	\label{silica}\\
	\rho&=&2.2\ \frac{\mbox{g}}{\mbox{cm}^3}, \ \
	C=6.7\times  10^6\ \frac{\mbox{erg}}{\mbox{g K}},\ \
	m=1.1\times 10^4\ \mbox{g},\ \nonumber \\
	E&=&7.2\times 10^{11}\frac{\mbox{erg}}{\mbox{cm}^3},  \ \
	\sigma=0.17,\ \
	\phi=5\times10^{-8}; \nonumber \\
\mbox{Sapphire:}
	\quad \alpha&=&5.0\times  10^{-6}\ \mbox{K}^{-1}, \ \
	\lambda^*=4.0\times  10^6\ \frac{\mbox{erg}}{\mbox{cm s K}},
	\label{saphire}\\
	\rho&=&4.0\ \frac{\mbox{g}}{\mbox{cm}^3},\ \
	C=7.9\times  10^6\ \frac{\mbox{erg}}{\mbox{g K}},\ \
	m=3\times 10^4 \ \mbox{g},\nonumber\\
	E&=&4\times 10^{12}\,\frac{\mbox{erg}}{\mbox{cm}^3},\ \
	\sigma=0.29,\ \
	\phi=3\times10^{-9}.
\end{eqnarray}

From these two figures it is evident that
the effects associated with thermal expansion in no case may be neglected. Especially unfortunate result
is obtained for sapphire -- the noise from thermodynamical fluctuations of
temperature dominates for frequencies of interest and is several times
larger than the SQL. While for fused silica the situation is the
opposite. The SQL is one order larger than TD fluctuations. It is important
to note, that unlike structural damping for which angles of losses at low
frequencies were not yet directly measured, our mechanism has fundamental
nature and is calculated explicitely without model assumptions.

\section*{Conclusion}
The analysis and numerical estimates presented above for two effects allow
to give some recommendations for the strategy of upgrading laser
interferometric gravitational wave antennae.

1. It is important to emphasize that contrastingly to the gravitational wave,
both effects do not displace the centers of masses of the mirrors. Thus,
in principle it is possible to subtract these effects as well as usual
Brownian fluctuations (with certain level of accuracy) in a way similar to
the procedure suggested for the subtraction of thermal noise in suspension
fibers \cite{blv}. However, till now to our knowledge nobody presented the scheme of
such subtraction.

2. Potentially there exists a possibility to decrease substantially these
effects by choosing materials for the mirrors with substantially smaller values
of $\alpha$. There also exists the possibility to use special cuts of
anisotropic monocrystals for which $\alpha$ turns to zero.

3. Thermodynamic fluctuations decreese with growing radius of the beam spot
as $r_0^{3/2}$ even stronger then for the Brownian fluctuations. In this way,
some optimization of the geometry of the resonators may help.

4. The "brute force" method for the improvement of the sensitivity by
increasing the circulating optical power $W$ does not look promising,
taking into account photo-thermal shot noise. It seems that few
mega-Watts is the upper limit for $W$.

\section*{Addendum}
The evident method to suppress the thermodynamical fluctuations by
using larger beamspots is unfortunately limited by diffraction losses and
technology. If we choose the LIGO-I geometry with beam sizes on internal
and end mirrors: $r_{0i}=3.6/\sqrt{2}\ \mbox{cm},\quad 
r_{0e}=4.6/\sqrt{2}\  \mbox{cm}$ 
then the threshold of sensitivity of thermodynamical fluctuations will be
only 1.7 times lower than on our graphics still 2 times higher than the SQL.

\section*{Acknowledgments}
We are pleased to thank for helpful discussions V.P.Mitrofanov, F.Ya.Khalili,
A.I.Osipov, I.A.Kvasnikov, H.-J.Kimble and R.Weiss.
This research was partially supported by the California Institute of
Technology, US National Science Foundation and the Russian Foundation for
Basic Research grant \#96-15-96780.

\section*{Appendix A.}

In this Appendix we calculate the spectral density (\ref{SDSN}) of the
fluctuations of the averaged displacement $\bar X$  caused by
photo-thermal shot noise.

We start from thermal conductivity  equation:
\begin{eqnarray}
\label{TCSN}
\frac{\partial  u}{\partial  t}-a^2\Delta u &=&
	\frac{w(t)}{\rho C}\frac{1}{\pi r_0^2}\ 2\delta(x)\
	\exp \left(-\frac{y^2+z^2}{ r_0^2} \right),\\
\langle w(t) w(t')\rangle&=& \hbar \omega_0 W_0 \delta(t-t'), \quad
	w(t)=W_{\rm abs}(t) - W_0, \quad
	W_0=\langle W_{\rm abs}\rangle \nonumber,
\end{eqnarray}
with boundary condition (\ref{boundaryTC}) and  equation of elasticity for the
field of deformation $\vec v(x,y,z)$ (see \cite{landau,boley}) with boundary conditions
for tensions on free surface:
\begin {eqnarray}
\label{EDfull}
&&\frac{1-\sigma}{1+\sigma}\, \mbox{grad div } \vec v -
	\frac{1-2\sigma}{2(1+\sigma)} \mbox{rot rot } \vec v =
	\alpha\, \mbox{grad } u, \\
\label{EDboundary}
&&\sigma_{xx}=\left.\frac{E}{1-2\sigma}\left[
	\frac{\sigma}{1+\sigma}\left(
		\frac{\partial v_x}{\partial x}+
		\frac{\partial v_y}{\partial y}+
		\frac{\partial v_z}{\partial z}\right) - \alpha u +
	\frac{1-2\sigma}{1+\sigma}\frac{\partial v_x}{\partial x} \right]
	\right|_{x=0}=0, 	\\
&&\sigma_{xy}=\left.\frac{E}{2(1+\sigma)}\left(
		\frac{\partial v_x}{\partial y}+
		\frac{\partial v_y}{\partial x}\right)
			\right|_{x=0}=0, \qquad
	\sigma_{xz}=\left.\frac{E}{2(1+\sigma)}\left(
		\frac{\partial v_x}{\partial z}+
		\frac{\partial v_z}{\partial x}\right)\right|_{x=0}=0,
\end{eqnarray}

The problem (\ref{TCSN}, \ref{boundaryTC}) in half-space can be replaced by
the problem (\ref{TCSN}) in full space, in which the function of source
is evenly continued for negative $x$. In this case the condition
(\ref{boundaryTC}) is satisfied  automatically.
Assuming that the absorbed power changes as
$w(t)=w(\omega)e^{i\omega t}$ one
can find the spectral component:
\begin{eqnarray*}
u(\vec r, \omega)&=&\int\!\int\!\int_{-\infty}^{\infty}\frac{dk_x dk_y dk_z}{(2\pi)^3}\,
	e^{ik_x x+ik_y y +ik_z z}\,e^{-r_0^2(k_y^2+k_z^2)/4}\,
        \frac{2w(\omega)}{\rho C(a^2k^2+ i\omega)}.
\end{eqnarray*}

After substitution of $u(\vec r, \omega)$ into (\ref {EDfull},
\ref{EDboundary})  one can find the solution for $\vec v$ as a sum \cite{boley}
$\vec v = \mbox{grad } \varphi +\vec v^{(a)}$, where $\varphi$ satisfies
Poisson's equation (it can be deduced from (\ref{EDfull})):
\begin{eqnarray}  \label{varphiPua}
\Delta \varphi =\frac{1+\sigma}{1-\sigma} \alpha u
\end{eqnarray}
without boundary conditions and function $ \vec v^{(a)}$ satisfies equation
(\ref {EDfull}) with zero right part and boundary conditions:
\begin{eqnarray}
\sigma_{xx}&=&\left. \frac{E}{1+\sigma}\left[
		\frac{\partial^2 \varphi}{\partial y^2}+
		\frac{\partial^2 \varphi}{\partial y^z} \right]
	\right|_{x=0},\label {boundaryva}\\
\sigma_{xy}&=-&\left.\frac{E}{(1+\sigma)}\,
		\frac{\partial^2 \varphi}{\partial x \partial y}
		\right|_{x=0}, 	\qquad
\sigma_{xz}=-\left.\frac{E}{(1+\sigma)}\,
		\frac{\partial^2 \varphi}{\partial x \partial z}
        \right|_{x=0}. 	\nonumber
\end{eqnarray}

It is convenient to choose the function $\varphi$ in symmetrical form
(by evenly continuing the $u(\vec r, \omega)$ for negative values of $x$):
\begin{eqnarray}
\label{simvarphi}
\varphi (x,y,z,\omega)&=&-\frac{\alpha (1+\sigma)}{4\pi (1-\sigma)}
	\int\!\int\!\int_{-\infty}^{\infty}
	\frac{dx'dy'dz'\,u(\vec r, \omega)}
	{\sqrt{(x')^2 + (y-y')^2 +(z-z')^2}}, \\
\varphi (x=0,y,z,\omega)&=&-\frac{\alpha (1+\sigma)}{(1-\sigma)}\,\frac{2w(\omega)}{\rho C}\,
	 \int\!\int\!\int_{-\infty}^{\infty}
        \frac{dk_x dk_y dk_z}{(2\pi)^3\, k^2(a^2k^2+i\omega)}\,
        e^{-r_0^2(k_y^2+k_z^2)/4 + ik_y y+ik_z z}\nonumber.
\end{eqnarray}
In this case $\partial_x \varphi(x=0,y,z)=0$ and therefore (i)
the function $\varphi$ gives no contribution into $\bar X$
 and (ii) tangential tensions $\sigma_{xy}$ and  $\sigma_{xz}$
in boundary conditions (\ref{boundaryva}) turn to zero.

The solution of the problem for $\vec v^{(a)}$ is known (see for example
\cite{landau}).  Using condition
(\ref{condition}) and  following table integrals:
\begin{eqnarray}
\label{int2}
\int\!\int_{-\infty}^{\infty}dy\,dz\,
        \frac{e^{ik_y y+ik_z z } }{\sqrt{y^2+z^2}}&=&
	\frac{2\pi}{k_\bot}, \quad k_\bot=\sqrt{k_y^2+k_z^2}.
\\
\label{int3}
\int\!\int\!\int_{-\infty}^{\infty}dx\,dy\,dz \,
	\frac {e^{ik_x x+ik_y y+ik_z z}} {\sqrt{x^2 + y^2 +z^2}}
        &=&\frac{4\pi}{k^2}, \quad k=\sqrt{k_x^2+k_y^2+k_z^2},
\end{eqnarray}
we substitute  $\vec
v^{(a)}$ into (\ref{barX}) and obtain:

\begin{eqnarray}
\bar X(\omega)&=&  - 4\alpha (1+\sigma)\,
	\frac{w(\omega)}{\rho C}
	\int\!\int\!\int_{\infty}^{\infty}
        \frac{dk_x dk_y dk_z}{(2\pi)^3}\,\frac{k_{\bot}}{k^2(a^2k^2+i\omega) }\,
	e^{-r_0^2(k_y^2+k_z^2)/2}\simeq\nonumber\\
&\simeq& - \frac{\alpha (1+\sigma)}{i\pi^2\omega}\,
	\frac{w(\omega)}{\rho C}
	\int_{-\infty}^{\infty}dk_x\int_{0}^{\infty} dk_{\bot}\,
        \frac{k_\bot}{k_x^2+k_\bot^2}\,
	e^{-r_0^2k_{\bot}^2/2}=
	- \frac{\alpha (1+\sigma)}{i\pi}\,
	\frac{w(\omega)}{\rho C r_0^2}\, \frac{1}{\omega}.
\end{eqnarray}
Assuming that spectral density of absorbed power is equal to
$S_w(\omega)=2 \hbar \omega_0 W_0$  one can find the formula (\ref{SDSN}).

\section*{Appendix B}

In this Appendix we deduce the formula (\ref{SDTD}) using Langevin abroach
for thermal conductivity equation (\ref{TDFT}) and recalculate the
thermo-dynamical fluctuations of temperature into deformations.

Using time and space Fourier transform one can write down the formal
solution of (\ref{TDFT}):
\begin{eqnarray} \label{u}
u(\vec r, t)&=&\int\!\int^{\infty}_{-\infty}\frac{d\vec k\, d\omega}{(2\pi)^4}\
        \frac{F(\vec k,\omega)}{a^2(\vec k )^2+ i\omega}
	e^{i\omega t +i\vec k\vec r},\label{ur}
\end{eqnarray}
We interest in temperature $\bar u$ averaged over some volume $V$. Assuming
that $\langle\,F(\vec k,\omega)F^*(\vec k',\omega')\, \rangle=(2\pi)^4\,
F_0^2|\vec k|^2 \delta (\vec k-\vec k')\, \delta (\omega-\omega')$ one can
calculate $\langle\, \bar u^2\, \rangle$
\begin{eqnarray}
\langle\, \bar u^2\, \rangle &=&
        \frac{1}{V} \int_V d\vec r   \frac{1}{V}\int_V d\vec r_1
        \int\!\int\!\int\!\int^{\infty}_{-\infty}
                \frac{d\vec k\, d\vec k'\, d\omega\, d\omega'}{(2\pi)^8}\
        \frac{\langle\,F(\vec k,\omega)F^*(\vec k',\omega')\,\rangle }
                {(a^2(\vec k )^2+ i\omega)(a^2(\vec k' )^2+ i\omega')}
	e^{i(\omega - \omega') t +i\vec k\vec r-i\vec k'\vec r_1}=\nonumber\\
 \label{baru2}
        &=&\frac {F_0^2}{2a^2}\, \frac{1}{V},\quad \rightarrow \quad
        F_0^2=2a^2\, \frac{\kappa T}{\rho C}.
\end{eqnarray}
Here we equate (\ref{baru2}) to (\ref{tvar}) and find $F_0^2$.

In order to find TD temperature fluctuations in half-space with boundary
condition (\ref{boundaryTC}) one can solve equivalent problem for full space
with source function even continued for negative $x$, so that its correlator
is equal to:
$$
\langle\,F(\vec k,\omega)F^*(\vec k',\omega')\, \rangle=
        (2\pi)^4\, F_0^2|\vec k|^2 \,
        [\delta (k_x-k_x')+\delta (k_x+k_x')]\,
        \delta (k_y-k_y')\,\delta (k_z-k_z')\,
        \delta (\omega-\omega')
$$

After that we must substitute the expression (\ref{u}) for $u$ into
elasticity equation
(\ref{EDfull}, \ref{EDboundary}). The solution  for $\vec v$  can be found
according to the same scheme as in Appendix A:
$\vec v = \mbox{grad } \varphi +\vec v^{(a)}$, where $\varphi$ satisfies
Poisson equation (\ref{varphiPua}). Again $\varphi$  can be written in
symmetrical form (\ref{simvarphi}) and the function $\vec v^{(a)}$ must satisfy
equation (\ref{EDfull}) with zero right part and following boundary conditions:
\begin{eqnarray*}
\sigma_{xx}(x=0,y,z)&=& \frac{\alpha E}{ (1-\sigma)}\,
        \int\!\int\!\int\!\int_{-\infty}^{\infty}
	\frac{dk_x dk_y dk_z d\omega}{(2\pi)^4}\,
        \frac{F(\vec k,\omega)}{a^2 k^2+ i\omega}\,
        \frac{k_y^2+k_z^2}{k^2}\,e^{i\omega t+ik_y y+ik_z z} ,\\
\sigma_{xy}(x=0,y,z)&=&\sigma_{xz}(x=0,y,z)=0.
 \end{eqnarray*}
The solution for $\vec v^{(a)}$ is known \cite{landau}
$$
v^{(a)}_x(x=0,y,z)= \frac{1-\sigma^2}{\pi\, E}\int\!\int_{-\infty}^{\infty}dy'dz'
\frac{\sigma_{xx}(x=0,y',z')}{\sqrt{(y-y')^2+(z-z')^2}}.
$$
Substituting it into (\ref{barX}) we find
$\bar X(t)$  and spectral
density $S_{TD}(\omega)$:
\begin{eqnarray*}
\bar X(t)&=&    2\alpha (1+\sigma)\,
        \int\!\int\!\int\!\int_{-\infty}^{\infty}
	\frac{dk_x dk_y dk_z d\omega}{(2\pi)^4}\,
        \frac{F(\vec k,\omega)}{a^2 k^2+ i\omega}\,
        \frac{k_\bot}{k^2 }\,
	e^{i\omega t-k_\bot^2r_0^2/4},\\
 S_{TD}(\omega)&=&
	8\alpha^2 (1+\sigma)^2\,
        \int\!\int\!\int_{-\infty}^{\infty}
	\frac{dk_x dk_y dk_z }{(2\pi)^3}\,
        \frac{F_0^2 }{a^4 k^4+ \omega^2}\,
        \frac{2 (k_\bot^2)}{k^2 }\,
	e^{-k_\bot^2r_0^2/2}
\end{eqnarray*}
From the last formula, using integration in spherical coordinate system
and neglecting the term $a^4 k^4$ in denominator according to condition
(\ref{condition}), we obtain formula (\ref{SDTD}) asymptotically correct
for the frequencies $\omega r^2_0/a^2\gg 1$.

The table integrals (\ref{int2}, \ref{int3}) was used above in calculations of
$\sigma_{xx}(x=0,y,z), \, v^{(a)}_x(x=0,y,z), \, \bar X(t)$.

\section*{Appendix C}
To calculate the fluctuations of the surface using FDT approach with thermoelastic
losses we should solve the system:
\begin{eqnarray}
\label{newv}
&&\frac{1-\sigma}{1+\sigma}\, \mbox{grad div } \vec v -
	\frac{1-2\sigma}{2(1+\sigma)} \mbox{rot rot } \vec v =
	\alpha\, \mbox{grad } u, \\
\label{newu}
&&\frac{\partial  u}{\partial  t}-a^2\Delta u =
	\frac{\alpha ET}{C\rho (1-2\sigma)}\frac{\partial\  {\rm div}\ \vec v}{\partial t}, \\
&& \sigma_{xx}=\frac{F_0}{\pi r_0^2}e^{-(y^2+z^2)/r_0^2} \quad \sigma_{xy}=\sigma_{xz}=0 \quad \left.{\frac{\partial u}{\partial x}}\right|_{x=0}=0.
\end{eqnarray}
In zero order approximation we simply solve elasticity problem neglecting
temperature $u$ in the right part of the first equation in (\ref{newv}) and
in boarder conditions. This solution is well known \cite{landau}:
\begin{eqnarray}
&&\vec v=\frac{F_0(1+\sigma)}{2\pi^2Er_0^2}\int\!\int_{-\infty}^\infty
	e^{-(y'^2+z'^2)/r_0^2}\vec G(\vec r-\vec r') dy' dz', \nonumber\\
&&\vec G(\vec r)=\left(\frac{2(1-\sigma)}{r}+\frac{x^2}{r^3}\right)\vec e_x+
	\left(\frac{x}{r^3}-
	\frac{1-2\sigma}{r(r+x)}\right)(y\  \vec e_y+ z\  \vec e_z).
\end{eqnarray}
By substituting this solution in (\ref{barX}) and using (\ref{int2}) we
obtain susceptibility $\chi^{(0)}$ (\ref{rigidity}). This result totally coincides
with the calculations for the half-space in \cite{french} (if we note that
the stored static strain energy $U=\bar X^{(0)}F_0/2=F_0^2\chi^{(0)}/2$ and
$w_0=\sqrt{2} r_0$) and has minor numerical difference with the approximation
in \cite{levin}.
To find first order approximation for the distribution of temperature we
substitute
\begin{equation}
{\rm div}\ \vec v^{(0)}=-\frac{(1+\sigma)(1-2\sigma)F_0}{2\pi^2 E}\int\!\int_{-\infty}^{\infty}
e^{-k_\bot^2 r_0^2/4-k_\bot x+ik_y y+ik_z z} dk_y \ dk_z
\end{equation}
in (\ref{newu}) to find that
\begin{equation}
u^{(1)}(\vec r,t)=\frac{4\alpha T(1+\sigma)F_0 i\omega }{C\rho}\int\!\int\!\int^\infty_{-\infty}\frac{dk_x\ dk_y\ dk_z}{(2\pi)^3}
	\frac{k_\bot}{k^2(a^2 k^2+i\omega)}e^{-k_\bot^2 r_0^2/4+i\vec k\vec r}.
\end{equation}
We then substitute this solution in (\ref{newv}) to find second order approximation
in the same way as in the two Appendices above by finding appropriate evenly
continued $\varphi$ function satisfying (\ref{varphiPua}, \ref{boundaryva}).
We find using (\ref{int3})
\begin{equation}
\sigma_{xx}(0,y,z)=-\frac{4(1+\sigma)\alpha^2TEF_0 i\omega}{(1-\sigma)\rho C}
\int\!\int\!\int_{-\infty}^\infty\frac{dk_x\ dk_y\ dk_z}{(2\pi)^3}\frac{k^3_\bot}
{k^4 (a^2k^2+i\omega)} e^{-k_\bot^2 r_0^2/4+ik_y y+ik_z z}.
\end{equation}
By substituting this expression in
\begin{equation}
v_x^{(2a)}(0,y,z)=\frac{1-\sigma^2}{\pi E} \int\!\int_{-\infty}^\infty dy'dz'
\frac{\sigma_{xx}(0,y',z')}{\sqrt{(y-y')^2+(z-z')^2}},
\end{equation}
and using (\ref{barX}) we may finally obtain
\begin{equation}
X^{(2)}=-8(1+\sigma)^2\frac{\alpha^2T}{\rho C} F_0 i\omega\int\!\int\!\int_{-\infty}^\infty
\frac{dk_x\ dk_y\ dk_z}{(2\pi)^3}\frac{k^2_\bot}{k^4(a^2 k^2+i\omega)}
e^{-k^2_\bot r_0^2/2}
\end{equation}
and hence ${\rm Im}(\chi(\omega))$ which leads through (\ref{spectrum}) to
precisely the same expression as (\ref{SDTD}).

\vfill\eject

\begin{figure}
\epsfbox{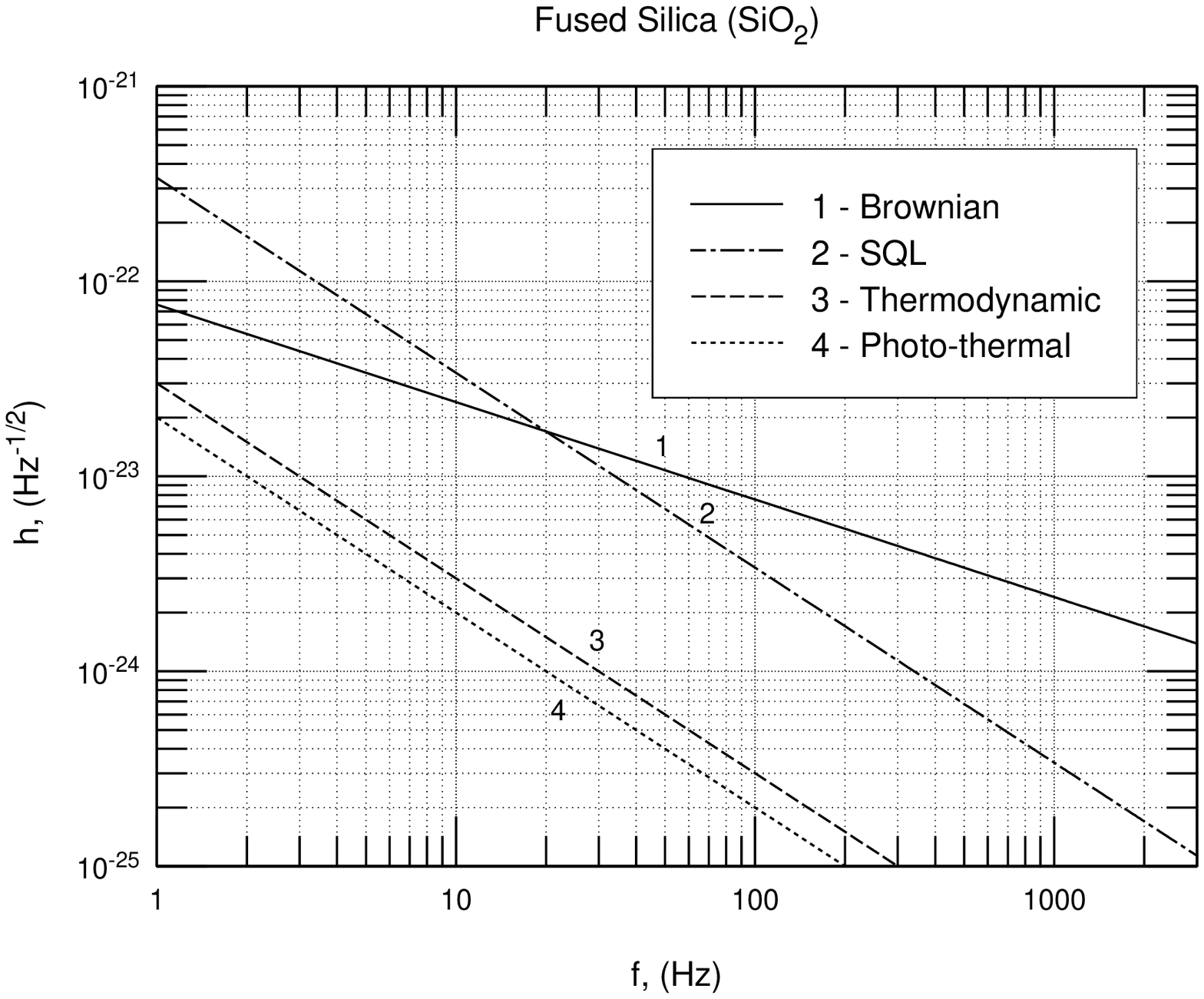}
\label{fig1}
\end{figure}

\vfill\eject

\begin{figure}
\epsfbox{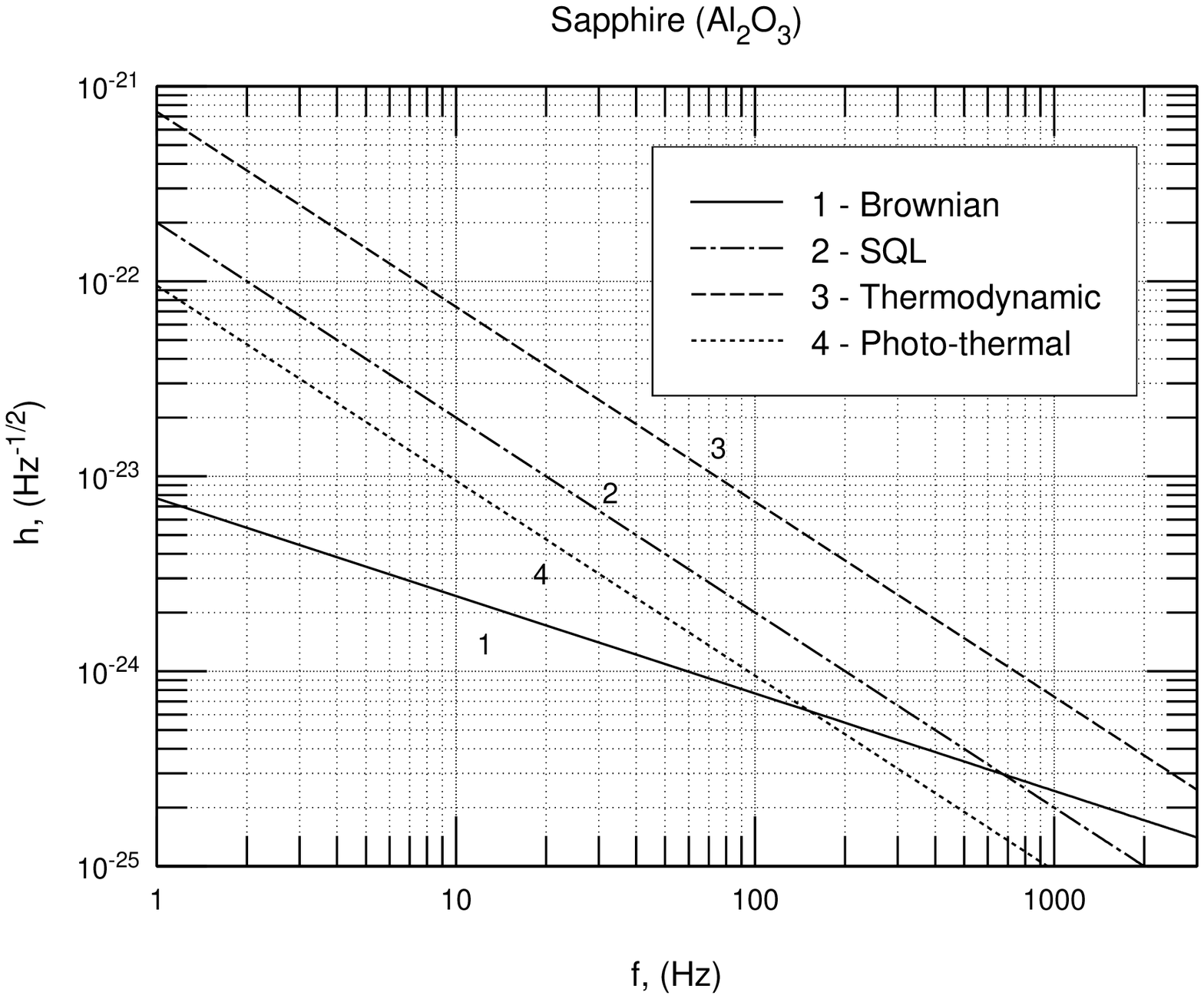}
\label{fig2}
\end{figure}

\vfill\eject

Figure captions:

\bigskip 
Fig.1 The sensitivity limitations for the perturbation of metric
caused by the SQL and by thermal fluctuations in fused silica mirrors.

\bigskip 
Fig.2  The sensitivity limitations for the perturbation of metric
caused by the SQL and by thermal fluctuations in sapphire mirrors.

\end{document}